# UAV Detection: A STDP trained Deep Convolutional Spiking Neural Network Retina-Neuromorphic Approach [?]

Paul Kirkland[1][0000 0001 5905 6816], Gaetano Di Caterina[1], John Soraghan[1], Yiannis Andreopoulos[2], and George Matich[3]

[1] The University of Strathclyde, Glasgow, United Kingdom
[2] University College London, London, United Kingdom
[3] Leonardo, London, United Kingdom

**Abstract.** The Dynamic Vision Sensor (DVS) has many attributes, such as sub-millisecond response time along with a good low light dy-namic range, that allows it to be well suited to the task for UAV De-tection. This paper proposes a system that exploits the features of an event camera solely for UAV detection while combining it with a Spik-ing Neural Network (SNN) trained using the unsupervised approach of Spike Time-Dependent Plasticity (STDP), to create an asynchronous, low power system with low computational overhead. Utilising the unique features of both the sensor and the network, this result in a system that is robust to a wide variety in lighting conditions, has a high temporal resolution, propagates only the minimal amount of information through the network, while training using the equivalent of 43,000 images. The network returns a 92% detection rate when shown other objects and can detect a UAV with less than 1% of pixels on the sensor being used for processing.

## 1 Introduction

Consumer UAVs and micro-UAVs are increasingly available at low cost, allowing their use in commercial applications (inspection, lming and deliveries)[14] and social use by the general public to become more frequent [19]. However, as the number of UAVs in circulation increases, so does the concern for misuse and accidents. A prime example in the UK recently was the closure of airports due to UAV ying over this restricted area [1], with near misses recorded in the UK in 2018 as 117, up 10 times from 4 years ago [5]. Nonetheless, a number of other concerns, other than collisions, exist due to the UAVs ability to carry a small payload: these could contain potentially harmful chemical or explosives or could be used to smuggle illegal goods [6].

[?] Supported by Leonardo, Data Collection in collaboration with University College London.

Detection of UAVs is not trivial due to their small form factor, coupled with the expanse of the search space. They also possess a high range of manoeuvrability while being di cult to discriminate against birds at distant ranges. These features make it di cult for typical detection approaches such as visual, infra-red, audio and radar to detect the UAV in a wide range of situations [6].

This paper presents a novel UAV Detection system, utilising the features of both the Dynamic Vision Sensor (DVS) and Spiking Neural Network (SNN). This end-to-end spiking Neuromorphic system possesses the following range of features: asynchronous functionality, low power consumption, low computational throughput, high dynamic range, high temporal resolution and dynamic relation-ship with scene environment. The results of a pilot study show that this system is ideal for the task of UAV detection, displaying features that are unmatched by any other single sensor systems.

The remainder of the paper is organised as follows. Section 2 provides back-ground on the sensor and the spiking network used and explaining the unsu-pervised learning mechanism. Section 3 provides details about the experimental set-up, Section 4 shows o the results of the system and Section 5 has the dis-cussion of these results.

## 2   Background

Neuromorphic engineering combines research from both the neuroscience and computational neuroscience elds that is exploited within an Engineering as-pect. The proposed system makes use of three such Neuromorphic approaches, the event-based camera, spiking neural network and spike time dependent plas-ticity. The respective sensor, neuron model and learning mechanism combine with a traditional Deep Convolutional Neural Network (DCNN) architecture, to capitalise on the characteristic unique to each.

### 2.1   Dynamic Vision Sensor - Event Based Camera

The Dynamic Vision Sensor is a biologically-inspired sensor (silicon retina) created to mimic how human eye perceives motion with their retina: as such the sensor asynchronously transmits the logarithmic light intensity di erence (events) on a pixel by pixel level. This replaces the xed frame rate traditional camera images, with a far more compressed and sparse output, resulting in 1 to 3 orders of magnitude increase in output rate (33 ms traditional to 15 s Event Based)
[4]. This allows the sensor to have a much higher temporal resolution (in essence a 66000 frames per second super slow-motion camera for up to 800 pixels, as compared to real world frames per second closer to 1-2,000) but without the caveat of the extra processing required for the pixels that didn't change. An-other feature is the DVS's high dynamic range, rated at >120dB vs the <60dB of traditional cameras [4, 11]. This allows the event based camera to see in a wide variety of lighting conditions, from quickly changing brightness conditions,

to low light ones, where traditional cameras would not be able to detect anything. A comparison of images captured from a DSLR and the DVS, showing UAVs ying in a well lit and low light scene, are illustrated in Figure 1. It can be seen that the DVS camera is able to capture the shape of the UAV in a well lit situation Fig.1(b) and (c) and is also able to capture the shape in the low light situation when the outline of the UAV is indistinguishable in Fig.1(e) and (f). The images in Fig.1 (c) and (f) show a typical post processing median ltering of the images to give better sensor noise suppression.

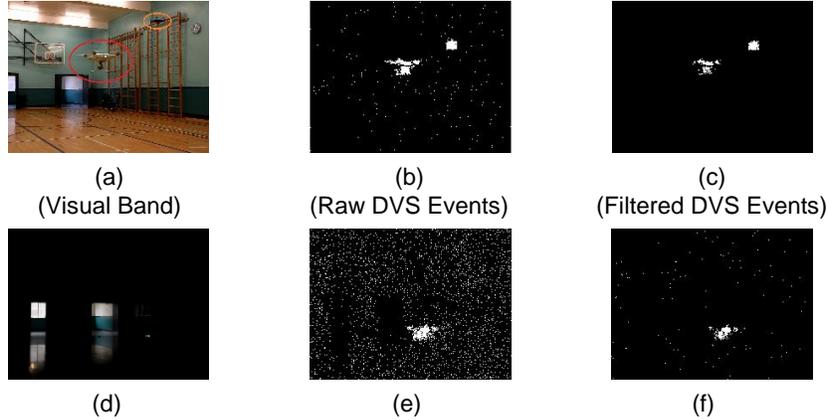

(a) (Visual Band)     (b) (Raw DVS Events)     (c) (Filtered DVS Events)

(d)     (e)     (f)

Fig. 1. Use of the High Dynamic Range within the DVS to capture stark lighting di erences. (TOP) Indoor well lit scene (BOTTOM) Low light scene.

The advantages of the DVS leads to the main attribute exploited within this paper, that relates the dynamic relationship to the visual source. This attribute is how the sensor can deliver a sparse yet detailed account of the scene, minimising computation and power. An example of this is shown in Figure 2 highlights the ability to change the integration time of the events captured to create a frame (for visual representation and training). The top row shows a slow-moving UAV, where a higher integration time is required to collect enough event to represent the UAV, as not as many changes in light intensity occur. While the bottom row illustrates the removal of motion blur, in a fast-moving UAV collision, by decreasing the integration time. The integration times can also be overlapped allowing a combination of both a longer integration time to capture events and the ne temporal resolution changes in the scene. The main drawback to the current DVS technology is the low spatial resolution. However, active research in this area has shown cameras with a sensor size of 640x480 [16] and 384x320 [9] pixels can be produced while maintaining the useful features.

### 2.2 Spiking Neural Network

The network used within this paper makes use of both the bene ts of convolutional and spiking neural networks.

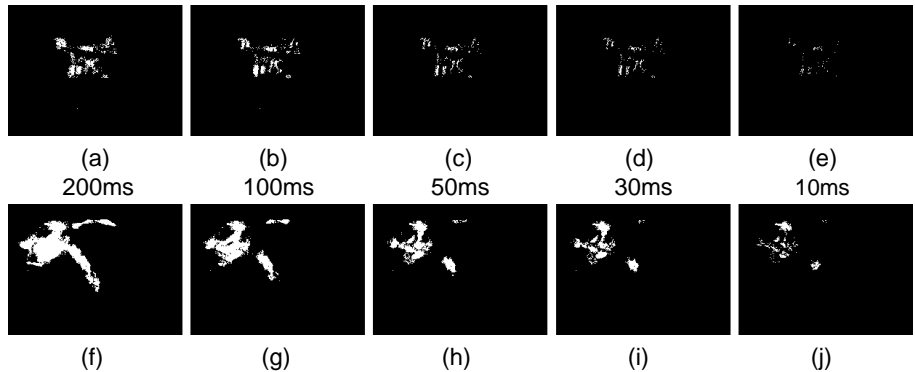

| (a) | (b) | (c) | (d) | (e) |
| 200ms | 100ms | 50ms | 30ms | 10ms |

| (f) | (g) | (h) | (i) | (j) |

Fig. 2. DVS ltered events captured in a range of time frames for a low speed (TOP)and high speed scene (BOTTOM)

The CNN brings a local spatial coherence and parameter/weight sharing method, that allows an image to be compressed, such that it can be represented by a respectively smaller number of features versus the number of pixel in the image. Combining these features within progressive layers allows fur-ther compression to occur. The SNN allows sparsity to occur through changing of the neuron model, and learning mechanism. The new neuron model con-verts the oating point values travelling through the network with 1 bit bi-nary spikes. These spikes are far more simplistic in nature, with constant am-plitude and duration of individual spikes. Their information is characterised entirely by their emission time (when a neuron red), and frequency of r-ing (how often a neuron res). The neurons have a threshold to reach be-fore passing information forward, but further information can be inferred from the timing of frequency of the neuron ring. In other words, only passing a small amount of important information through the network, but in a timely manner. This can be seen as similar to that of the primate visual system, which has been shown to have spike rates on the order of a few hertz [15]. The change in the neuron model leads to a very important paradigm shift in the network: from looking for con-tent, to looking for context. This reit-erates the usefulness of the sparse in-formation transfer that can relate im-portance into it's time dependency. In that, a few import pieces of context can be used to build content, but no amount of content can give you con-text. This type of sparse, spike-time-based deep network [10, 12, 17] is not as suited for a backpropagation learning mechanism as a CNN. It then makes use of a simpli ed unsupervised Spike-Time Dependent Plasticity (STDP) rule [2] in combination with a winner-takes-all (WTA) approach, to extract hierarchical features in CNN-like architecture. The described network, illustrated in Figure 3, shows the typical three convolution layers with pooling layer in-between. Unlike CNN learning, with STDP each convolution layer has an intra and inter lateral inhibition mechanism [18, 10]. This helps the network to reduce the information propagated, especially redundant and repeating information, while ensuring that the most salient information is maintained. It operates by only allowing one fea-ture (neuron) in a feature (neuron) map to re per frame, seen as an intra map competition. This WTA approach then moves onto the inter map inhibition. Only allowing one spike to occur in any given spatial region, typically the size of the convolution kernel, throughout all the maps. When not training the con-volution and pooling layers operate in a standard procedure, with the pooling also following the WTA theme with a max pool operation.

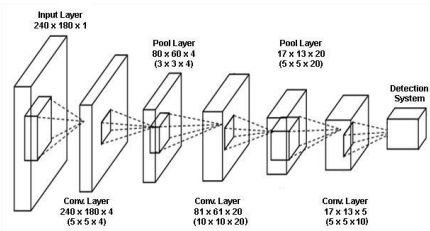

Fig. 3. SNN Architecture 3 Layer Convolution and Pooling

## 3    Methodology

Two di erent methods of capturing data were used in this work. Actual events captured from a DVS and simulated events generated from data captured from a higher resolution DSLR camera. The following describes the set up for the experimental data collection and simulated sensor data creation along with any speci cs pertaining to the network and learning mechanism.

### 3.1    Dynamic Vision Sensor Data

The data was captured within a small (basketball court sized) gymnasium, as seen in Figure 4, in order to be able to control the amount of light in a given scene. Two di erent sized UAVs (DJI Phantom [without propellers - 290

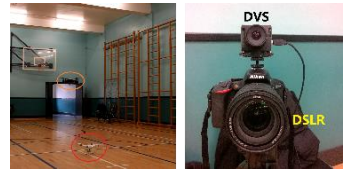

Fig. 4. Indoor Test Set Up

x 290 x 195 mm] and DJI Tello [98 x 92.5 x 41 mm]) were used which allowed a wider range of test scenarios to be replicated. The event data was captured using a DVS240 Neuromorphic Vision Sensor with a spatial resolution of 240x180 and asynchronous event output. It was mounted on top of a DSLR Camera, produc-ing a 1920x1080 output at 60 Frames per Second (FPS), both pictured in Figure 4. This was used for ground truth data and use within the simulated data as a means of comparison. The DVS camera is set up to give out a tuple for each spike event, these contain the xy coordinate, the timestamp of when the event occurred, and the polarity of the change in intensity. However, during training and testing of the proposed system the polarity value was ignored, compress-ing all the spike information into one channel instead of two. The time-stamp data was embedded into each of the frames used for the dataset, this provides a signi cant advantage over simulated event data [10] as the earlier events are no longer just the highest contrast, but actually, still represent the spatio-temporal domain they were captured in. To further improve this temporal aspect a range of integration time for the dataset frame collection was used, ranging from 10ms up to 200ms with overlaps in the time windows of 10%, 50% and 90%. This

6     P. Kirkland et al.

wide variety in the frames allows the sensor to capture a diverse range of speed variability within the sensor eld of view. This allows the temporal data, usually lost in the snapshot of a frame to be instilled within event capture.

### 3.2 Simulated Dynamic Vision Sensor Data

There are two sources of UAV footage used for the simulated event data: Video recorded from a DSLR camera as explained in the previous section. The other footage is captured from some outside testing using a DSLR(1920x1080 @ 30-60fps).

An example of the simulated data is provided in Figure 5, which shows a simulated events frame, along with the pre and post the processing stage when the resolution is down-sampled to 240x180. The outdoor footage provides a wide range of lighting conditions per frame and a number of di er-ent background disturbances to cause noise and clutter in the data (clouds and ground objects). The only issue with simulating the event data is the inher-ent lack of temporal resolution (missing information between each frame that needs to be interpolated). A number of simulators already exist [3, 7, 13], with PIX2NVS [3] being the event simulator used in this paper. The event sensor sim-ulator takes the frame rate and interpolates the events that would exist between the frames. This is limited to the actual recorded frame rate of the footage, so to further enhance the temporal resolution, some extra post processing is carried out to reduce the number of spikes. Allowing a higher delity capture of only the edges of the moving objects.

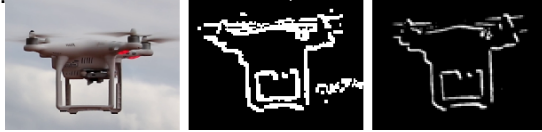

Fig. 5. Simulated UAVs

Table 1. Table of Network Parameters.

| Layer | Conv. 1 | Pool 1 | Conv. 2 | Pool 2 | Conv. 3 |
|---|---|---|---|---|---|
| Filter Size | 5 | 5 | 10 | 5 | 5 |
| Number of Maps | 4 | 4 | 20 | 20 | 10 |
| Stride | 1 | 5 | 1 | 5 | 1 |
| Propagation Threshold | / Input Spikes | 1 | 45 | 1 | 3 |
| Initial Weights | Mean of 0.8 with STD 0.08 | | | | |

### 3.3 Proposed UAV Detection System

As mentioned in Section 3.1 the asynchronous data produced by the DVS cam-era is converted into a frame, embedded with the temporal data (a image where the value of a pixel is the time-stamp of the event occurrence). This frame is then used within a layer-wise learning methodology to extract and build fea-tures to allow the network to successfully identify a UAV. A list of the network parameters is shown in Table 1, which also highlights a novel feature of the proposed system, pre-emptive neuron thresholding (PENT). The PENT takes the typically reactive neuron thresholding concept [8], but allows it to work in advance of the spikes reaching a neuron. This concept is to overcome the potential of spikes saturating the rst layer of the SNN causing a false detection in the system. A typical reactive system would adapt the neuron thresholds if the saturation continued over time, but with the PENT approach, the system is able to act in a timely manner to prevent such saturation from propagating false features through the network. The detection parameters for nding a UAV are embedded within the network itself. As the network enforces a WTA approach to convolution and pooling, the last convolution layer as seen in Figure 3, has a highly sparse input and output. This allows it to act as a detection layer. In this situation, it is able to forgo usage of a fully connected layer [18] or support vector machine [10] as classi cation isn't required. The network's evaluation will be based upon the number of successful detections and its robustness to a range of highly spiking noisy inputs, replicating low light conditions. The proposed system will use data captured from both the actual DVS and the simulated DVS. The aim is to show how a network can be trained to deliver a higher ac-curacy from extending training data with simulated DVS data. This data would often be easier to obtain or would already exist, highlighting the ease at which a traditional visual detection system could be converted to an event camera and SNN. With this conversion resulting in a notable reduction in computational, processing and power, promoting its use within an environment where resources are limited.

## 4    Testing Results

This section shows the results of training from three UAV detection networks, using only actual DVS event data, only simulated DVS event data, and our proposed system which utilises both of the previous datasets together. Each of the networks is then tested on a series of actual DVS Event frames, comparing the bene t of additional training data, even if it is simulated data.

DVS Trained Network - During initial testing the net-work trained on real DVS event data struggled to converge to useful feature within the second layer, due to the sparse feature maps that were learned in the rst layer. A set of pre-trained weights representing Gabor features, shown in Figure 6, indicative of that seen in other rst layer SNNs [10, 18], allowed all the networks to have better building blocks to create more complex features in the second and third layers. Throughout all of the further testing, this method was used, using the four features presented in Fig-ure 6 as the rst layer of each network. Training the network using the events captured from the DVS results in a low resolution feature combination for the second layer. These feature maps resemble low delity UAV shapes as seen in Figure 7. It also shows the progression of these shapes into the third layer used for detection. This network produced an overall accuracy of 90% when using PENT (50-54% with static thresholding depending on the focus of true or false positives). Results are located in Table 2, which shows the overall accuracy along with the confusion matrix for each of the following networks.

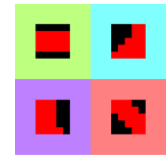

Fig. 6. Synthetic Gabor Features

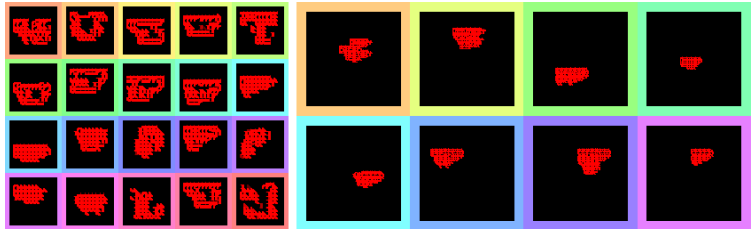

Fig. 7. Second and Third Layer of the Actual DVS Event Trained Network

Table 2. Results data in the confusion matrix for the three trained networks.

| N = 2000 | | | | UAV Predicted | | | | |
|---|---|---|---|---|---|---|---|---|
| | | True | False | | True | False | | True | False |
| Actual UAV | True | 850 | 150 | True | 610 | 490 | True | 880 | 120 |
| | False | 50 | 950 | False | 40 | 960 | False | 60 | 940 |
| Overall Accuracy | Actual DVS = 90% | | | Simulated DVS = 78% | | | Proposed System = 91% | | |

Simulated DVS Trained Network - The network trained using simulated DVS event data was then tested for comparison. These simulated events are originating from a higher resolution image then being scaled to the same resolution as the actual DVS. As the scene is derived from a higher resolution, a higher delity feature can occur in the second layer, as seen in Figure 8. These higher delity features combine with the low, to create features more representative of a UAV in the third layer. This seemingly qualitative improvement results in a quantitative drop in overall accuracy down to 78% with full results in Table 2. The drop in accuracy is a result of the features of the network having to ne a delity compared to the actual DVS test set. However, this network did re-turn the best false positive results, suggesting these more complex features were better at discriminating objects in the images without UAVs.

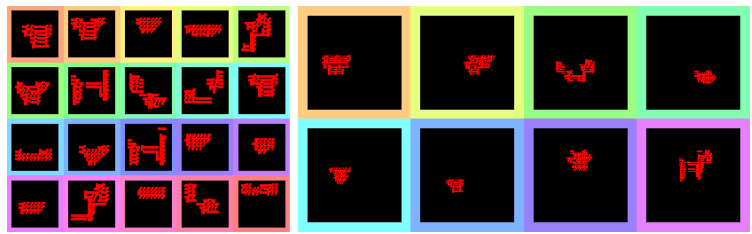

Fig. 8. Second and Third Layer of the Simulated DVS Event Trained Network

Proposed System Trained Network - The proposed detection system is the third network to be trained, utilising both datasets, real and simulated. At first this network exhibits visually very similar features in the second and third layers as seen in Fig. 9(d) and (h) to that of the simulation data network shown in Figure 8. Figure 9 also demonstrates how the network learns the features seen in these layers, started with the random weight Fig.9(a) and (e), then refining the important features in UAV shaped component parts seen in Fig.9(b),(c),(f) and
(g). While the proposed system and previous network trained on simulated data appear to have learned the same feature mapping, the accuracy results show otherwise with an overall accuracy of 92% exhibiting the highest number of correct detection, results shown in Table 2. To help visualise how these features help to detect the UAV, an example of an event image from both the actual and simulated DVS data is shown in Figure 10, indicating where the pooled mapping of the features map onto the UAV. The image also highlights how an improvement in spatial resolution of the sensor could open up the possibilities of UAV classification system rather than just detection systems. By using the higher fidelity features from the extra spatial resolution, it allows a better realisation on the component part of the UAV allowing more distinct feature to exist.

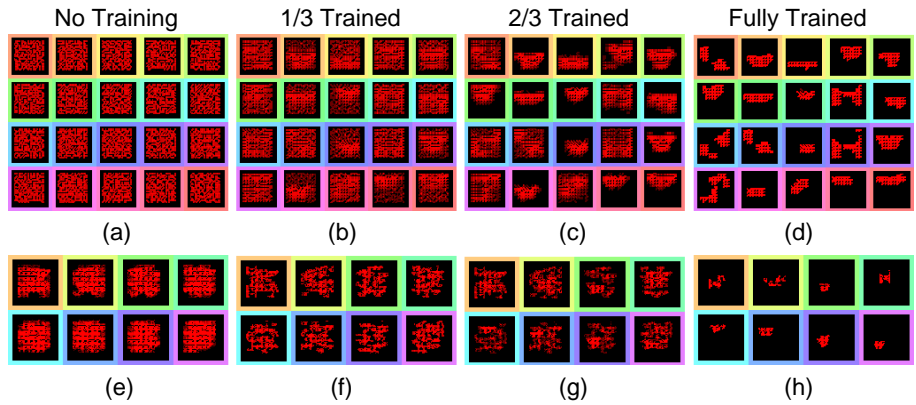

Fig. 9. Illustration of training in the UAV Detection Network

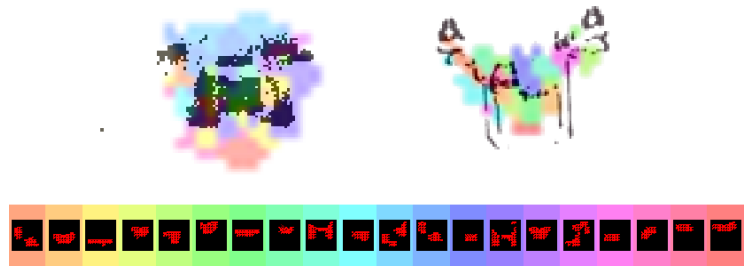

Fig. 10. Feature Mapping and their activations on UAV image

This demonstrates the main contribution of this paper that simulated DVS events can be a useful training tool for the desired network, when used in conjunction with actual DVS event data, improving upon the network trained only on DVS events. Since traditional video is more regularly available, this can prove an excellent starting point for new ideas and concepts that might not have DVS event footage. This could be ideal in situations where new data is either difficult to generate or obtain.

The proposed system was also able to show robustness to noise with the second contribution of this paper being the introduction of PENT (Pre-Emptive Neuron Thresholding). A visualisation of how noise is handled by the PENT is shown in Figure 11, depicting events captured from a low light scene with a UAV flying, similar to that shown in the low light scene Fig.1 (d). Demonstrating how when PENT is active, only the features of the UAV are captured as shown in Fig.11(b), while when PENT is off, the UAV features are masked by noise seen in Fig.11(f). The reduced propagation of features through the network due to saturation of the first layer has an impact on all subsequent layers as shown in Fig.11(c),(d),(g) and (h). This illustrates the difference PENT makes on the overall network. Quantitative testing was carried out on a range of artificial noise levels: 23dB to -16dB by adding 0.5% to 5% pixels worth of noise to the image respectively, results are displayed in Table 3. The results show that even when there are more noise event pixels active than event pixels that represent the UAV, the system can still determine the presence of a UAV.

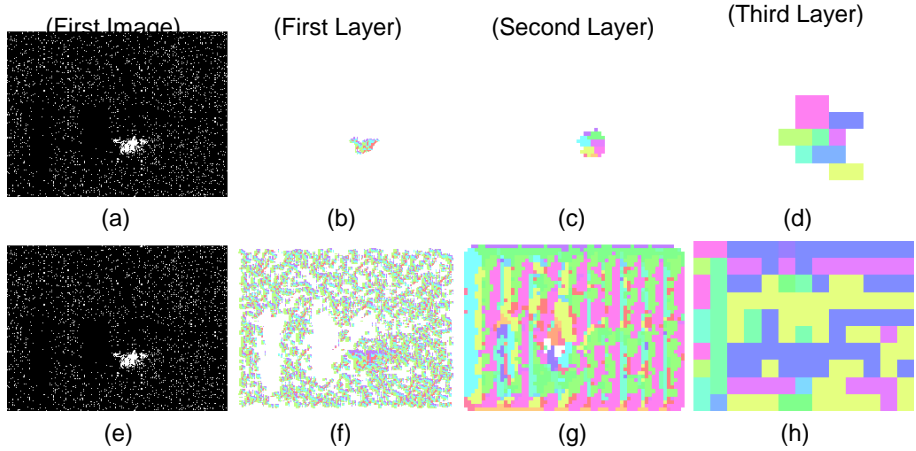

Fig. 11. Active threshold stops the saturation of layer one features propagation through the network causing false detections.

Table 3. Table of Accuracy with Additive Noise.

| SNR Level (dB) | 23 | 15 | 7 | 0.1 | -9 | -15 |
|---|---|---|---|---|---|---|
| Accuracy (%) | 85 | 83 | 82 | 72 | 62 | 46 |

The results from the SNN resemble those from a CNN, thanks partially to the convolution and pooling layers. Furthermore, the system is built upon a sparse spiking neuron model which only further sparsifies throughout the network, while in an unsupervised fashion learns distinctive feature to identify a UAV. This sparsity instils the ethos of only transmitting important information, which results in a lower computational throughput, for both runtime and training. The training sees a further reduction in information transfer due to the STDP allowing the system to converge to useful feature quickly. This results in using only 20,000 images each to train the second and third layer (3000 required if you want to train layer one), so 40-43,000 in total. Overall the spiking CNN would have a large reduction in computation need compared to its equivalent CNN, with this reduction in computation being a considerable factor in many applications.

## 5 Conclusion

Consumer UAVs and micro-UAVs have presented security and defence with a new-age problem. This paper presents a robust detection system for UAVs, that has many of the useful features of other sensors, while fewer of the drawbacks. The overall accuracy of 92%, coupled with an enhanced resilience to noise due to PENT, make the proposed system a feasible alternative for the future. From utilising the sparse nature of the SNN, this accuracy comes with the benefit of also providing a far lower computation load than a traditional CNNs, this being a result of not having to pass information from every neuron in the layer, but only those who pass the threshold. The SNN also pairs nicely with the asynchronous event driven nature of the DVS. With its output also representing a sparse version of the traditional frame based camera. The system to that effect then delivers high accuracy, while being the sparse version of the traditional system. This sparsity can deliver many benefits with reductions in computational processing leading to a reduction in overall size, weight, power and cost, therefore improving overall application system viability.